\documentclass[12pt]{book}

\usepackage[dvips]{graphicx,color}
\usepackage{makeidx,tsukuba}

\makeauthorindex
\makeindex

\begin{document}

\BookTitle{\itshape The 28th International Cosmic Ray Conference}
\CopyRight{\copyright 2003 by Universal Academy Press, Inc.}
\pagenumbering{arabic}

\chapter{Radio Emission From EAS - Coherent Geosynchrotron Radiation}

\author{Tim Huege,$^1$ and Heino Falcke$^{1,2}$ \\
{\it (1) Max-Planck-Institut f\"ur Radioastronomie, Auf dem H\"ugel 69, 53121 Bonn, Germany\\
(2) Adjunct Professor, Dept.\ of Astronomy, University of Nijmegen, P.O.\ Box 9010, 6500 GL Nijmegen, The Netherlands}}

\section*{Abstract}

Extensive air showers (EAS) have been known for over 30 years to emit pulses of radio emission at frequencies from a few to a few hundred MHz, an effect that offers great opportunities for the study of EAS with the next generation of ``software radio interferometers'' such as LOFAR and LOPES. The details of the emission mechanism, however, remain rather uncertain to date. Following past suggestions that the bulk of the emission is of geomagnetic origin, we model the radio pulses as ``coherent geosynchrotron radiation'' arising from the deflection of electrons and positrons in the earth's magnetic field. We analytically develop our model in a step-by-step procedure to disentangle the coherence effects arising from different scales present in the shower structure and infer which shower characteristics govern the frequency spectrum and radial dependence of the emission. The effect is unavoidable and our predictions are in good agreement with the available experimental data within their large margins of error.

\section{Introduction}

Research concerning radio emission from EAS almost ceased completely in the 1970s. With the advent of fully digital ``software radio interferometers'' such as LOFAR, however, it once more becomes very attractive to measure EAS through their radio emission. The technique is also ideally suited to be combined with classical particle detector arrays such as KASCADE Grande (as evident from the LOPES experiment described in [3]) or AUGER and offers the advantage of a very high duty cycle and cost-effectiveness over air fluorescence methods.

Past experiments have established that the dominant emission mechanism is related to the earth's magnetic field. The strength of the emission is, however, still rather unclear and past modeling efforts have not been developed to sufficient depth for application to concrete experiments. We therefore take the new approach of modeling the radio emission as ``coherent geosynchrotron emission'' as first proposed by [2] and developed in depth in [4] and [8]. Compared with recent Monte-Carlo simulations by [8] we take a more analytical approach in hope to better understand the various coherence effects shaping the emission and additionally take into account the longitudinal shower evolution.

\section{The Model}

Starting with the works of [5] we base our model on the frequency spectra of synchrotron pulses emitted by highly relativistic electron-positron pairs:

\begin{equation}\label{eqn:Epairpart}
\vec{E}(\vec{R},\omega) = \left(\frac{4 \pi}{c}\right)^{1/2} \frac{1}{R}\ \frac{\omega e}{\sqrt{8 c}\pi} \mathrm{e}^{\mathrm{i} (\omega \frac{R}{c}-\frac{\pi}{2})}  \left(-\vec{\hat{e}}_{\parallel}\right) A_{\parallel}(\omega),
\end{equation}
where
\begin{equation}
A_{\parallel}(\omega)=\mathrm{i} \frac{2 \rho}{\sqrt{3} c} \left(\frac{1}{\gamma^{2}}+\theta^{2}\right) K_{2/3}(\xi)
\end{equation}
with
\begin{equation}
\xi = \frac{\omega \rho}{3 c} \left(\frac{1}{\gamma^{2}}+\theta^{2}\right)^{3/2} \quad \mathrm{and} \quad \rho = \frac{\gamma m_{e} c^2}{e B \sin \alpha}.
\end{equation}

Our gradual development of the integration over the shower particles helps in understanding the coherence effects arising from the different scales present in the shower. Examination of a point source demonstrates the importance of the intrinsic beaming of the synchrotron pulses for the radial dependence of the emission. Going from a point source to a longitudinal line charge of a few metres length (corresponding to the thickness of the shower ``pancake'') shows that the associated coherence losses limit the frequency spectrum to the regime $<100$~MHz. Extending the model to a spherical shell and later on a ``flaring disk'' reveals coherence effects arising from a more realistic shower geometry. We take the lateral particle distribution into account via NKG-functions and incorporate the longitudinal spread as a function of distance from the shower axis through empirical data. An enhancement of the emission at high distances in the E-W compared to the N-S direction arises from the dependence on the geomagnetic field. In the last step we integrate over the longitudinal shower evolution as a whole, which has a strong impact on the total emission strength and the radial dependence of the emission. We also consider the energy distribution of the particles.

\section{Comparison with Data}

At the moment, the available historic data on radio emission from EAS is subject to large uncertainties. This is mainly due to discrepancies in the emission strength of an order of magnitude between the data sets from different groups, probably arising from radio calibration issues, and a lack of precision in the documentation of the more than 30 years old data sets. Additionally, there are some subtleties involved in the conversion of theoretical values ($E_{\omega}$) to measured quantities ($\epsilon_{\nu}$). We have chosen the well documented data of [1] and compared them with our predictions for the radial dependence of the emission of a $10^{17}$~eV vertical air shower in Figure 1. For an assessment of the spectral dependence as shown in Figure 2 we have taken data from [6] and [7] and {\em{rescaled}} them so that they are consistent with the data of [1].

\begin{figure}[t]
  \begin{center}
    \includegraphics[height=14.3pc]{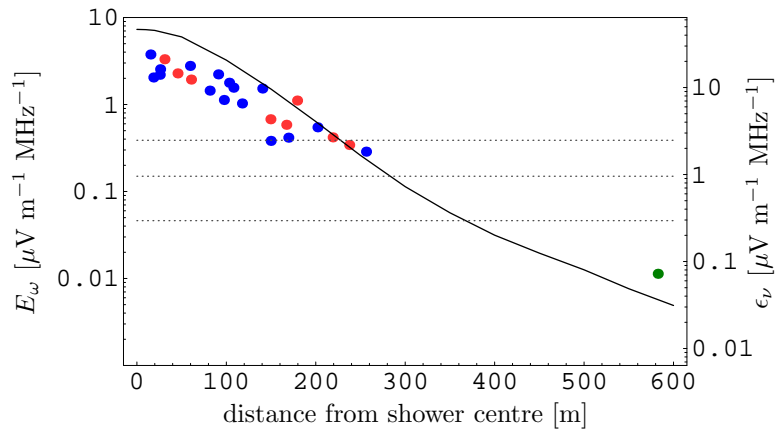}
  \end{center}
  \vspace{-2.0pc}
  \caption{Radial dependence (N-S direction) of the 55~MHz emission from a $10^{17}$~eV air shower with shower maximum at 4~km height. Data from [1]. Horizontal lines represent 3$\sigma$-detection for a LOPES station with 1/10/100 antennas. E-W direction would be significantly enhanced at high distances. For details see [4].}
\end{figure}

\begin{figure}[t]
  \begin{center}
    \includegraphics[height=14.3pc]{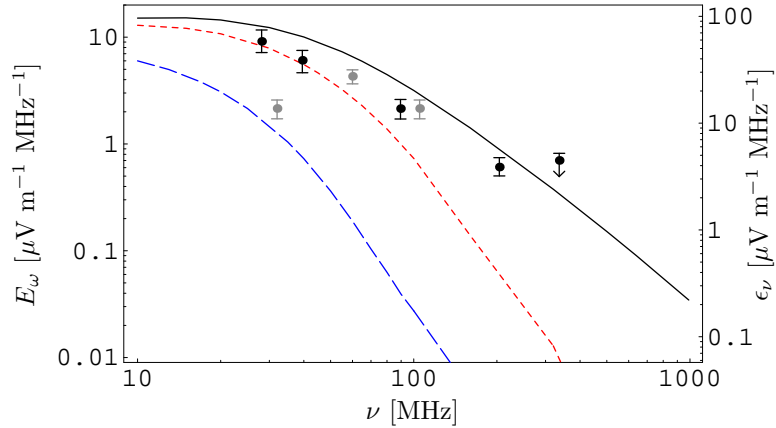}
  \end{center}
  \vspace{-2.0pc}
  \caption{Spectral dependence for same shower as in Fig.\ 1. Solid: centre of illuminated area; short-dashed/long-dashed: 100~m/250~m from centre in N-S direction; grey/black points: {\em{rescaled}} data from [6]/[7]. For details see [4].}
\end{figure}

\section{Discussion}

As can be seen in Figures 1 and 2, our calculations successfully reproduce the trends visible in the data sets. As for the absolute strength of the emission, our model somewhat over-predicts the data of [1], which is already at the upper end of the available data sets. Given the very simplified integration over the longitudinal shower evolution this is, however, not too surprising. We also find a significant E-W to N-S anisotropy in the emission strength (not shown here). In general, it is thus clear that a significant contribution by the geomagnetic emission mechanism is unavoidable. At the moment, the acquisition of reliable experimental data is of paramount importance. As our signal-to-noise calculations indicate, this will be accomplished by LOPES in the near future, see [3].

\section{Conclusions}

Our model of ``coherent geosynchrotron radiation'' reproduces the available experimental data within their large margins of error. The step-by-step modeling procedure illustrates that the high-frequency cutoff in the emission spectra is mainly governed by the longitudinal spread, i.e.\ thickness, of the shower ``pancake''. The radial dependence is governed by the intrinsic beaming of the pulses as well as the integration over the shower evolution as a whole. The predicted asymmetry of emission strength in N-S and E-W directions will be verifiable with LOPES and would help in the detection of EAS out to high distances in the E-W direction. In a next step, we will use this model as a basis for the development of a sophisticated numerical simulation that includes additional aspects such as near-field effects, polarisation and Askaryan-type \v Cerenkov radiation.

\section{List of Symbols/Nomenclature}

  \begin{tabbing}
     A = BCDEFGHIJKLMNOPQRST                 \=       \kill
     $\vec{R}$ = distance observer-particle  \>  $\vec{E}$ = electric field \\
     $c$ = speed of light                    \>  $e$/$\mathrm{m}_{e}$ = electron unit charge/mass \\
     $\omega$ = 2$\pi$ observing frequency   \>  $\theta$ = minimum angle to line of sight \\
     $\gamma$ = particle Lorentz factor      \>  $K_{\nu}$ = modified Bessel-function of order $\nu$ \\
     $B$ = earth's magnetic field            \>  $\alpha$ = pitch angle of particle trajectory

  \end{tabbing}

\section{References}

\re
1.\ Allan H.~R., Clay R.~W., Jones J.~K.\ 1970, Nature 227, 1116
\re
2.\ Falcke H., Gorham P.~W.\ 2003, Astropart.\ Phys.\ in press, astro-ph/0207226
\re
3.\ Horneffer A., Falcke H.\ 2003, Proc.\ of the 28th ICRC, these proceedings
\re
4.\ Huege T., Falcke H.\ 2003, A\&A submitted
\re
5.\ Jackson J.~D.\ 1975, Classical Electrodynamics (J.\ Wiley \& Sons, New York)
\re
6.\ Prah, J.~H.\ 1971, M.Phil. thesis, University of London
\re
7.\ Spencer R.~E.\ 1969, Nature 222, 460
\re
8.\ Suprun D.~A., Gorham P.~W., Rosner J.~L.\ 2003, astro-ph/0211273

\endofpaper
\end{document}